\documentclass[preprintnumbers]{revtex4}
\UseRawInputEncoding
\usepackage [latin1]{inputenc}

\usepackage{amssymb}
\usepackage{amsmath}
\usepackage{graphicx}
\usepackage{dcolumn}
\usepackage{bm}
\usepackage{subfigure}
\usepackage{color}

\begin{document}

\title{Phase transition of non-linear charged Anti-de Sitter black holes}
\author{Yun-Zhi Du, Huai-Fan Li, Fang Liu, Ren Zhao, Li-Chun Zhang\footnote{the corresponding author}}
\affiliation{Department of Physics, Shan xi Da tong University, Da tong 037009, China\\
Institute of Theoretical Physics, Shan xi Da tong University, Da tong 037009, China}

\thanks{\emph{e-mail:zhlc2969@163.cn}}

\begin{abstract}
Understanding the thermodynamic phase transition of black holes can provide a deep insight into the fundamental properties of black hole gravity to establish the theory of quantum gravity. We investigate the condition and latent heat of phase transition for non-linear charged AdS black holes using the Maxwell's equal-area law, and analysis the boundary and curve of the two-phase coexistence area in the expanded phase space. We suggest that the phase transition of the non-linear charged AdS black hole with the fixed temperature ($T<T_c$) is related to the electric potential at the horizon, not only to the location of horizon. Recently, the molecular number density was introduced to study the phase transition and microstructure of black holes. On this basis, we discuss the continuous phase transition of a non-linear charged AdS black hole to reveal the potential microstructure of a black hole by introducing the order parameter and using the scalar curvature.
\end{abstract}

\maketitle

\section{Introduction}
Black hole physics, as an intersection of many disciplines between the GR and quantum mechanics, such as thermodynamic statistical physics and particle physics, plays a significant role in modern physics. The study of black holes from the perspective of thermodynamics has become  indisputable. Black holes have the characteristics of an ordinary thermodynamical system and they can undergo phase transitions under certain conditions. Black holes have thermal properties; hence, a key concern regarding black hole is that whether they have a microstructure like ordinary thermodynamic systems. From a mathematical perspective, string theory and supersymmetry are used to explain the microstructure of black holes. From a thermodynamic view point, hypotheses of black hole molecules and spacetime atoms are proposed to explore the microscopic behavior of black holes. Research on both areas has made progress \cite{Ruppeiner2014,Ruppeiner2018,Ruppeiner2008,Ruppeiner2012,Strominger1996,Emparan2006,Horowitz1996,Miao2018,Miao2019,Miao2017,Wei2015,Hendi2017}

By regarding the cosmological constant in the AdS spacetime as the pressure of an ordinary thermodynamic system and comparing the state parameters of the AdS black holes to that of the van der Waals (vdW) system, the thermodynamical phase transition continue to be one of the increasingly active areas in black hole physics. Some satisfactory results on the thermodynamical property of AdS and dS black holes were obtained \cite{Hendi2017a,Hennigar2017a,Frassin,Kubiznak2012,Cai2013,Ma2017,Ma2017a,Mir2017b,Banerjee2017,Banerjee2011,Hendi2019,Simovic2019,Hennigar2019,Mbarek2019,
Kubiznak2016,Li2017a,Li2017,zhao2015,Ma2018,Ma2016,Zhang2016,Zhou2019,Wei2019}. However, the statistical mechanics background of black holes as a thermodynamic system is not understood yet. Therefore, obtaining the relationship between the thermodynamic properties of various AdS and dS black holes is crucial. This study will be useful for further understanding the entropy, temperature, heat capacity, etc, of black holes and developing a self-consistent black hole thermodynamic geometry theory.

For the AdS black holes, discontinuous exist in the P-V or T-S diagram with $T<T_c$ or $P<P_c$ \cite{zhao2015,Ma2018}. The black holes within the two-phase coexistence zone have the latent heat of phase transition; they have different thermodynamical properties at different phases. Therefore, a black hole as a thermodynamical system must have different microstructures with different phases.

The linear charged black holes in AdS spacetime \cite{Chamblin1999,Chamblin1999a} within a second order phase transition shows a scaling symmetry: at the critical point, the state parameters scale with respect to charge q, i.e., $S\sim q^2,~P\sim q^{-2},~T\sim q^{-1}$ \cite{Johnson2018,Johnson2018a}. One can infer whether the scaling symmetry exists in the non-linear charged AdS black holes \cite{Hassaine2008}. As a generalization of the charged AdS Einstein-Maxwell black holes, exploring new non-linear charged systems is interesting. Due to infinite self-energy of point like charges in Maxwell's theory \cite{Born1934,Born1934a,Kats2007,Anninos2009,Cai2008,Seiberg1999,Fradkin1985,Mesaev1987,Bergshoeff1987,Tseytlin1986,Gross1987}, Born and Infeld proposed a generalization when the field is strong, introducing non-linearities \cite{Born1934,Born1934a,Dirac2013,Birula1970}. An interesting non-linear generalization of charged black holes involves a Yang-Mill field coupled to Einstein gravity, where several features in extended thermodynamics have recently been studied \cite{Zhang2015,Moumni2018}.

As a unique perspective, the thermodynamic geometry (scalar curvature) plays an important role in studying black hole phase transition. The comprehensible physical picture of Ruppeiner geometry makes it widely used \cite{Ruppeiner2014,Ruppeiner2018,Ruppeiner2008,Ruppeiner2012,Miao2019,Miao2017,Wei2015,Chamblin1999a,Johnson2018,Johnson2018a,Born1934,Born1934a,Kats2007}. In the case of fixed charge, considering the fluctuation of mass and pressure, we will study the Ruppeiner geometric line element and scalar curvature. We predict that for the non-linear charged AdS black hole with a certain temperature, the different electric potentials correspond to different values of the scalar curvature. In contrast, this result indicate different microstructures of black hole system with different electric potentials.

In this study, we will investigate the phase transition of the non-linear charged AdS black hole (EPYM black hole) to obtain the condition and the latent heat of phase transition by using Maxwell's equal-area law and adopting different independent dual parameters. We aim to reveal the microstructure of black holes by studying the phase transition of the EPYM black hole with different electric potentials. This work is organized as follows. In Sec. \ref{scheme2}, we present the thermodynamic parameters of the EPYM black hole. In Sec. \ref{scheme3}, we discuss the phase transition of the EPYM black hole for different choices of the independent dual parameters by the Maxwell¡¯s equal-area law and provide a universal condition of phase transition. In Sec. \ref{scheme4}, we present the coexistence curve in the P-T diagram and analyze the influence of non-linear YM charged parameter $\gamma$ on the latent heat of phase change for the EPYM AdS black hole. In Sec. \ref{scheme5}, the order parameter $\phi^2$ is introduced to explain the phase transition for the EPYM black hole. Moreover, the thermodynamic geometry at the critical point is analyzed using the scalar curvature $R$. We also investigate the influence of the non-linear YM charge parameter $\gamma$ on $R$. Finally, a brief summary is provided in Sec. \ref{scheme6}.

\section{Thermodynamics for Non-Linear Charged AdS Black Holes}
\label{scheme2}
Our starting point is the action for four-dimensional Einstein-power-Yang-Mills (EPYM) gravity with a cosmological constant $\Lambda$, given by \cite{Zhang2015,Lorenci2002,Corda2011,Mazharimousavi2009}
\begin{eqnarray}
I=\frac{1}{2}\int d^4x\sqrt{g}
\left(R-2\Lambda-[Tr(F^{(a)}_{{\mu\nu}}F^{{(a)\mu\nu}})]^\gamma\right)
\end{eqnarray}
with the Yang-Mills (YM) field
\begin{eqnarray}
F_{\mu \nu}^{(a)}=\partial_{\mu} A_{\nu}^{(a)}-\partial_{\nu} A_{\mu}^{(a)}+\frac{1}{2 \xi} C_{(b)(c)}^{(a)} A_{\mu}^{(b)} A_{\nu}^{(c)}.
\end{eqnarray}
Here, $Tr(F^{(a)}_{\mu\nu}F^{(a)\mu\nu})
=\sum^3_{a=1}F^{(a)}_{\mu\nu}F^{(a)\mu\nu}$, $R$ and $\gamma$ are the scalar curvature and a positive real parameter, respectively; $C_{(b)(c)}^{(a)}$ represents the structure constants of three parameter Lie group $G$; $\xi$ is the coupling constant; and $A_{\mu}^{(a)}$ represents the $So(3)$ gauge group YM potentials.

For this system, the four-dimensional EPYM black hole solution with the negative cosmological constant $\Lambda$ is obtained by adopting the metric \cite{Yerra2018}
\begin{eqnarray}
d s^{2}=-f(r) d t^{2}+f^{-1} d r^{2}+r^{2} d \Omega_{2}^{2},\\
f(r)=1-\frac{2 M}{r}-\frac{\Lambda}{3} r^{2}+\frac{\left(2 q^{2}\right)^{\gamma}}{2(4 \gamma-3) r^{4 \gamma-2}},
\end{eqnarray}
where $d\Omega_{2}^{2}$ is the metric on unit 2-sphere with volume $4\pi$ and $q$ is the YM charge. Note that this solution is valid for the condition of the non-linear YM charge parameter $\gamma\neq0.75$, and the power YM term holds the weak energy condition (WEC) for $\gamma>0$ \cite{Corda2011}. The position of the black hole event horizon is determined as the larger root of $f(r_+)=0$. The parameter $M$ represents the ADM mass of the black hole; in our set up, it is associated with the enthalpy of the system.

Using the ``Euclidean trick'', one can identify the black hole temperature and entropy of the solution given by \cite{Zhang2015}
\begin{eqnarray}
T=\frac{1}{4 \pi r_{+}}\left(1+8 \pi P r_{+}^{2}-\frac{\left(2 q^{2}\right)^{\gamma}}{2 r_{+}^{(4 \gamma-2)}}\right),~~~~~S=\pi r_{+}^{2}.    \label{T}
\end{eqnarray}
The pressure reads $P=-\Lambda /(8 \pi)$ from extended thermodynamics; the YM potential $\Psi$ is given by \cite{Anninos2009,Dehyadegari2017}
\begin{eqnarray}
\Psi=\frac{\partial M}{\partial q^{2 \gamma}}=\frac{r_{+}^{3-4 \gamma} 2^{\gamma-2}}{(4 \gamma-3)}.\label{Psi}
\end{eqnarray}
The above thermodynamic state parameters satisfy the first law
\begin{eqnarray}
d M=T d S+\Psi d q^{2 \gamma}+V d P
\end{eqnarray}
with the thermodynamic volume
\begin{eqnarray}
V=\left(\frac{\partial M}{\partial P}\right)_{S, q}=\frac{4 \pi}{3} r_{+}^{3}.\nonumber
\end{eqnarray}
Thus, the ADM mass of a black hole corresponds to enthalpy \cite{Kastor2009} as follows \cite{Ma2017a}:
\begin{eqnarray}
H=M(S, q, P)=\frac{1}{6}\left[8 \pi P\left(\frac{S}{\pi}\right)^{3 / 2}+3\left(\frac{S}{\pi}\right)^{\frac{3-4 \gamma}{2}} \frac{\left(2 q^{2}\right)^{\gamma}}{8 \gamma-6}+3 \sqrt{\frac{S}{\pi}}\right].
\end{eqnarray}
The equation of state $P(V,T)$ for canonical ensemble (fixed YM charge $q$) can be obtained from the expression of temperature as
\begin{eqnarray}
P=\left(\frac{4 \pi}{3 V}\right)^{1 / 3}\left[\frac{T}{2}-\frac{1}{8 \pi}\left(\frac{4 \pi}{3 V}\right)^{1 / 3}+\frac{\left(2 q^{2}\right)^{\gamma}}{16 \pi}\left(\frac{4 \pi}{3 V}\right)^{\frac{1-4\gamma}{3}}\right].\label{PTV}
\end{eqnarray}

\section{The Equal-Area Law}
\label{scheme3}

From Eq. (\ref{PTV}), we know that the state equation of the EPYM black hole with fixed YM charge corresponds to an ordinary thermodynamic system; it can be written as $f(T,P,V)=0$. Furthermore, the number of particles in the system is unchanged. In the following, through the Maxwell' equal-area law, we discuss the phase transition condition of the EPYM black hole in $P-V$, $T-S$, and $q^{2\gamma}-\Psi$, respectively. Then, we provide a general condition of the phase transition for the EPYM black hole.

\subsection{The construction of equal-area law in P-V diagram}

For the EPYM black hole with the given YM charge $q$ and temperature $T_0<T_c$, the volume at the boundary of the two-phase coexistence area are $V_1$ and $V_2$, respectively; the corresponding pressure $P_0$ is less than the critical pressure $P_c$; it is determined by the horizon radius $r_+$. Therefore, from the Maxwell's equal-area law $P_0(V_2-V_1)=\int^{V_2}_{V_1}PdV$ and Eq. (\ref{T}), we have
\begin{eqnarray}
P_{0}&=&\frac{T_{0}}{2 r_{1}}-\frac{1}{8 \pi r_{1}^{2}}+\frac{\left(2 q^{2}\right)^{\gamma}}{16 \pi r_{1}^{4 \gamma}},\nonumber\\
\quad P_{0}&=&\frac{T_{0}}{2 r_{2}}-\frac{1}{8 \pi r_{2}^{2}}+\frac{\left(2 q^{2}\right)^{\gamma}}{16 \pi r_{2}^{4 \gamma}},  \label{P0}\\
2 P_{0}&=&\frac{3 T_{0}(1+x)}{2 r_{2}\left(1+x+x^{2}\right)}-\frac{3}{4 \pi r_{2}^{2}\left(1+x+x^{2}\right)}\nonumber\\
&&+\frac{3\left(2 q^{2}\right)^{\gamma}\left(1-x^{3-4 \gamma}\right)}{8 \pi(3-4 \gamma) r_{2}^{4 \gamma}\left(1-x^{3}\right)} \label{PP0}
\end{eqnarray}
where $x=\frac{r_1}{r_2}$. With Eq. (\ref{P0}), we have
\begin{eqnarray}
0&=&T_{0}-\frac{1}{4 \pi r_{2} x}(1+x)+\frac{\left(2 q^{2}\right)^{\gamma}}{8 \pi r_{2}^{4 \gamma-1} x^{4 \gamma-1}} \frac{\left(1-x^{4 \gamma}\right)}{(1-x)}, \label{T0}\\
2 P_{0}&=&\frac{T_{0}}{2 r_{2} x}(1+x)-\frac{1}{8 \pi r_{2}^{2} x^{2}}\left(1+x^{2}\right)\nonumber\\
&&+\frac{\left(2 q^{2}\right)^{\gamma}}{16 \pi r_{2}^{4 \gamma} x^{4 \gamma}}\left(1+x^{4 \gamma}\right).\label{PPP0}
\end{eqnarray}
From the above equation and Eq. (\ref{PP0}), we obtain the following expression:
\begin{eqnarray}
\frac{1}{4 \pi r_{2} x}&=&T_{0} \frac{(1+x)}{\left(1+3 x+x^{2}\right)}\nonumber\\
&&+\frac{\left(2 q^{2}\right)^{\gamma}\left[(3-4 \gamma)\left(1-x^{3+4 \gamma}\right)+(3+4 \gamma) x^{3}\left(1-x^{4 \gamma-3}\right)\right]}{8 \pi(3-4 \gamma) r_{2}^{4 \gamma-1} x^{4 \gamma-1}(1-x^3)(1-x)^{2}\left(1+3 x+x^{2}\right)}.\label{T00}
\end{eqnarray}
Considering Eqs. (\ref{T0}) and (\ref{T00}), we have
\begin{eqnarray}
r_{2}^{4 \gamma-2}&=&\frac{\left(2 q^{2}\right)^{\gamma}\left[(3-4 \gamma)(1+x)\left(1-x^{4 \gamma}\right)+8 \gamma x^{2}\left(1-x^{4 \gamma-3}\right)\right]}{2 x^{4 \gamma-2}(3-4 \gamma)(1-x)^{3}}\nonumber\\
&=&\left(2 q^{2}\right)^{\gamma} f(x,\gamma).
\label{r2}
\end{eqnarray}
For the critical point ($x=1$), the state parameters are
\begin{eqnarray}
r_{c}^{4 \gamma-2}&=&\left(2 q^{2}\right)^{\gamma} f(1, \gamma), \quad f(1, \gamma)=\gamma(4 \gamma-1),\\
T_{c}&=&\frac{1}{\pi\left(2 q^{2}\right)^{\gamma /(4 \gamma-2)} f^{1 /(4 \gamma-2)}(1, \gamma)} \frac{2 \gamma-1}{4 \gamma-1}, \\
P_{c}&=&\frac{2 \gamma-1}{16 \pi \gamma\left(2 q^{2}\right)^{\gamma /(2 \gamma-1)} f^{1 /(2 \gamma-1)}(1, \gamma)}.\label{rcTcPc}
\end{eqnarray}
Because the above state parameters must be positive, the non-linear YM charge parameter satisfies the condition $\frac{1}{2}<\gamma$.

Substituting Eq. (\ref{r2}) into Eq. (\ref{T0}) and setting $T_0=\chi T_c$ ($0<\chi\leq1$), we have
\begin{eqnarray}
T_{0}&=&\frac{1}{4 \pi x\left(2 q^{2}\right)^{\gamma /(4 \gamma-2)} f^{1 / (4 \gamma-2 )}(x,\gamma)}\nonumber\\
&&\times\left(1+x-\frac{1}{2 f(x, \gamma) x^{4 \gamma-2}} \frac{1-x^{4 \gamma}}{1-x}\right),\label{T0final}
\end{eqnarray}
\begin{eqnarray}
&&\chi \frac{2 \gamma-1}{\gamma^{\frac{1}{4 \gamma-2}}(4 \gamma-1)^{\frac{4 \gamma-1 }{4 \gamma-2}}}\nonumber\\
&&=\frac{1}{4 x f^{1 /(4 \gamma-2)}(x, \gamma)}\left(1+x-\frac{1}{2 f(x, \gamma) x^{4 \gamma-2}} \frac{1-x^{4 \gamma}}{1-x}\right).\label{xchi}
\end{eqnarray}

\begin{figure}[htp]
\centering
\includegraphics[width=0.45\textwidth]{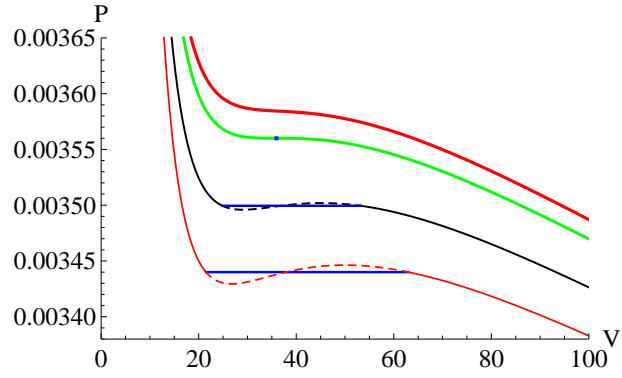}
\caption{Phase transition in $P-V$ diagram with the parameters $q=0.85,~\gamma=0.8$. The temperature is set to $0.0419$ (red thin line), $0.04215$ (black thin line), $0.0424$ (green thick line), and $0.0425$ (red thick line). }\label{P-V}
\end{figure}

\begin{figure}[htp]
\centering
\includegraphics[width=0.45\textwidth]{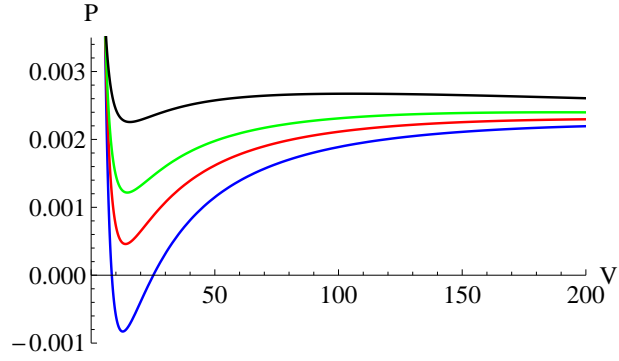}
\caption{Phase transition in $P-V$ diagram with the parameters $q=0.85,~T_0=0.0374$. The parameter $\gamma$ is set to 0.78 (black line), 0.9 (green line), 1 (red line), and 1.2 (blue line), respectively. }\label{P-V-gamma}
\end{figure}

For the given parameter $\gamma$ and temperature $T_0$ (i.e., $\chi$), we can obtain the value of $x$ from Eq. (\ref{xchi}). Eq. (\ref{r2}) suggests that for the given temperature $T_0$ ($T_0<T_c$), that is, for the fixed value of $x$, the phase transition condition reads
\begin{eqnarray}
\frac{\left(2 q^{2}\right)^{\gamma}}{r_{2}^{4 \gamma-2}}=\frac{1}{f(x, \gamma)}. \label{PTC}
\end{eqnarray}
Therefore, the phase transition of the EPYM black hole with a given temperature $T_0$ ($T_0<T_c$) is determined by the radio between the YM charge $(2q^2)^\gamma$ and $r_{2}^{4 \gamma-2}$ and not the value of the horizon alone. Note that we call this radio the YM electric potential with the horizon radius $r_2$. The phase transitions in the $P-V$ diagram with different temperatures are shown in Fig. \ref{P-V}. The effect of the non-linear parameter $\gamma$ on phase transition is exhibited in Fig. \ref{P-V-gamma}


\subsection{The construction of equal-area law in T-S diagram}

\begin{figure}[htp]
\centering
\includegraphics[width=0.45\textwidth]{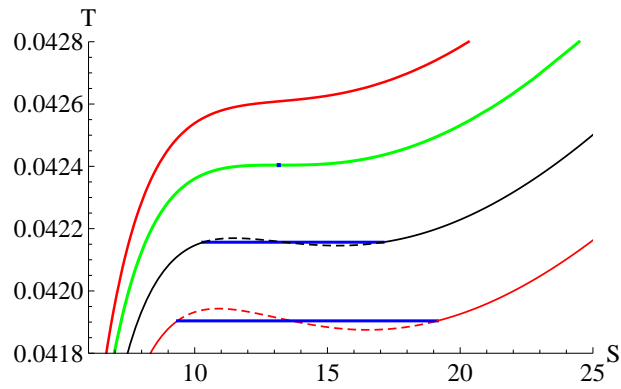}
\caption{Phase transition in $T-S$ diagram with the parameters $q=0.85,~\gamma=0.8$. The temperature is set to $0.000344$ (red thin line), $0.0035$ (black thin line), $0.0036$ (green thick line),  and $0.00365$ (red thick line).}\label{T-S}
\end{figure}

For the EPYM black hole with the given YM charge $q$ and pressure $P_0<P_c$, the entropy at the boundary of the two-phase coexistence area are $S_1$ and $S_2$, respectively. The corresponding temperature $T_0$ is less than the critical temperature $T_c$; it is determined by the horizon radius $r_+$. Therefore, from the Maxwell's equal-area law $T_0(S_2-S_1)=\int^{S_2}_{S_1}TdS$ and eq. (\ref{T}), we have
\begin{eqnarray}
2 \pi T_{0}&=&\frac{1}{r_{2}(1+x)}+\frac{8 \pi P r_{2}}{3(1+x)}\left(1+x+x^{2}\right)
\nonumber\\
&&-\frac{\left(2 q^{2}\right)^{\gamma} r_{2}^{1-4 \gamma}}{2(3-4 \gamma)} \frac{\left(1-x^{3-4 \gamma}\right)}{\left(1-x^{2}\right)},\\
T_{0}&=&\frac{1}{4 \pi r_{2}}\left(1+8 \pi P r_{2}^{2}-\frac{\left(2 q^{2}\right)^{\gamma}}{2 r_{2}^{(4 \gamma-2)}}\right),\nonumber\\
T_{0}&=&\frac{1}{4 \pi r_{1}}\left(1+8 \pi P r_{1}^{2}-\frac{\left(2 q^{2}\right)^{\gamma}}{2 r_{1}^{(4 \gamma-2)}}\right).
\end{eqnarray}
Considering the above equations, we obtain that the phase transition condition with the independent dual parameters for $T$-$S$ is the same as that for $P$-$V$. This result indicates that these two choices of independent dual parameters ($P$-$V$, $T$-$S$) will provide the same phase transition point. The curves of phase transition for $T$-$S$ are shown in Fig. \ref{T-S}.

\subsection{The construction of equal-area law in $q^{2\gamma}-\Psi$ diagram}
For the EPYM black hole with the given temperature $T_0<T_c$, if the YM charge changes, we will choose $q^{2\gamma}-\Psi$ to study the phase transition. The YM electric potential at the boundary of the two-phase coexistence area are $\Psi_1$ and $\Psi_2$, respectively. The corresponding YM charge term is $q^{2\gamma}_0$. Therefore, from the Maxwell's equal-area law $q^{2\gamma}_0(\Psi_2-\Psi_1)=\int^{\Psi_2}_{\Psi_1}q^{2\gamma}d\Psi$ and eq. (\ref{Psi}), we have
\begin{eqnarray}
q_{0}^{2 \gamma}&=&\frac{(4 \gamma-3)(1-x) x^{4 \gamma-3}}{2^{\gamma-1}\left(1-x^{4 \gamma-31}\right)} r_{2}^{4 \gamma-2}\nonumber\\
&&\times\left[1+\frac{8 \pi P_{0}}{3} r_{2}^{2}\left(1+x+x^{2}\right)-2 \pi T_{0} r_{2}(1+x)\right],\\
\frac{q_{0}^{2 \gamma}}{2^{1-\gamma}}&=&r_{2}^{(4 \gamma-2)}+8 \pi P_{0} r_{2}^{4 \gamma}-T_{0} 4 \pi r_{2}^{4 \gamma-1},\nonumber\\
\frac{q_{0}^{2 \gamma}}{2^{1-\gamma}}&=&r_{1}^{(4 \gamma-2)}+8 \pi P_{0} r_{1}^{4 \gamma}-T_{0} 4 \pi r_{1}^{4 \gamma-1}.
\end{eqnarray}
With the above equations, we have
\begin{eqnarray}
r_{2}^{4 \gamma-2}=\left(2 q_{0}^{2}\right)^{\gamma} \frac{(4 \gamma-3)(1-x)\left[(1+2 x)+x^{4 \gamma-1}(2+x)\right]-6 x^{2}\left(1-x^{4 \gamma-3}\right)}{2 x^{4 \gamma-2}(4 \gamma-3)(1-x)^{3}},
\end{eqnarray}
which is of the same form as Eq. (\ref{r2}).

\begin{figure}[htp]
\centering
\subfigure[$\gamma=0.8,~P_0=0.0034$]
{\includegraphics[width=0.45\textwidth]{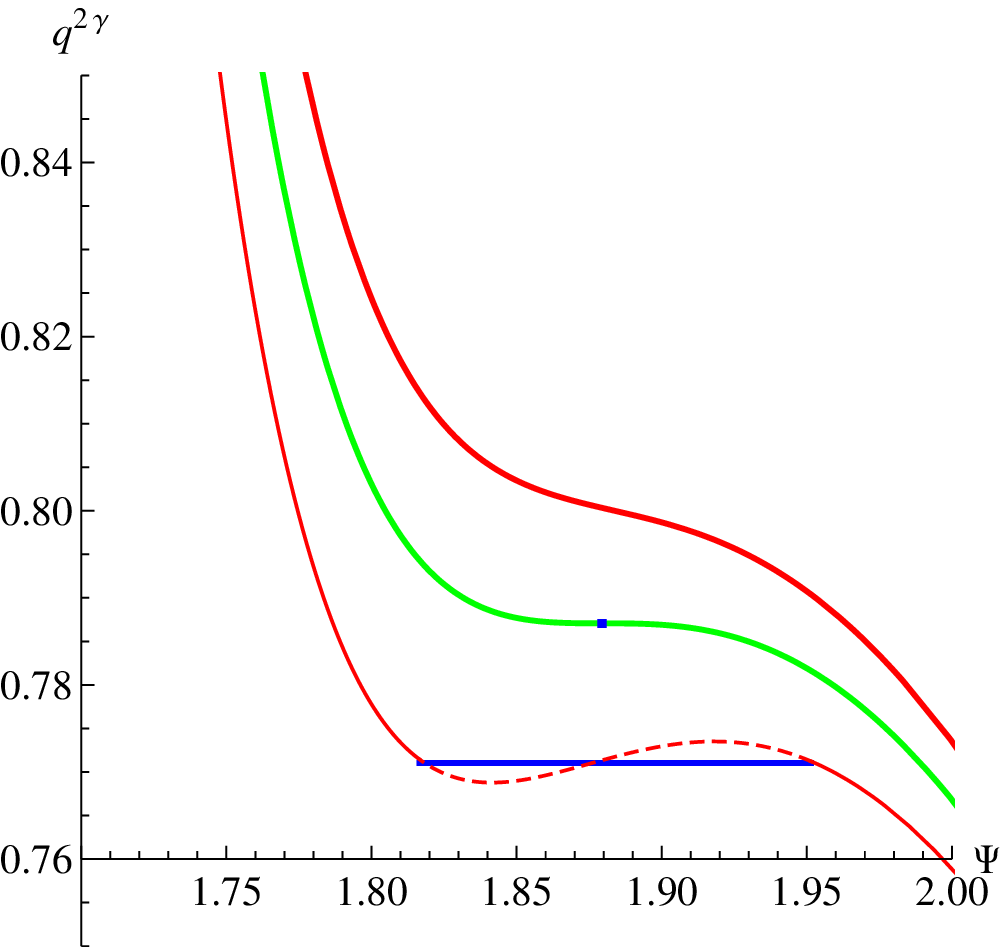}~~}
\subfigure[$\gamma=0.8,~T_0=0.0419$]
{\includegraphics[width=0.45\textwidth]{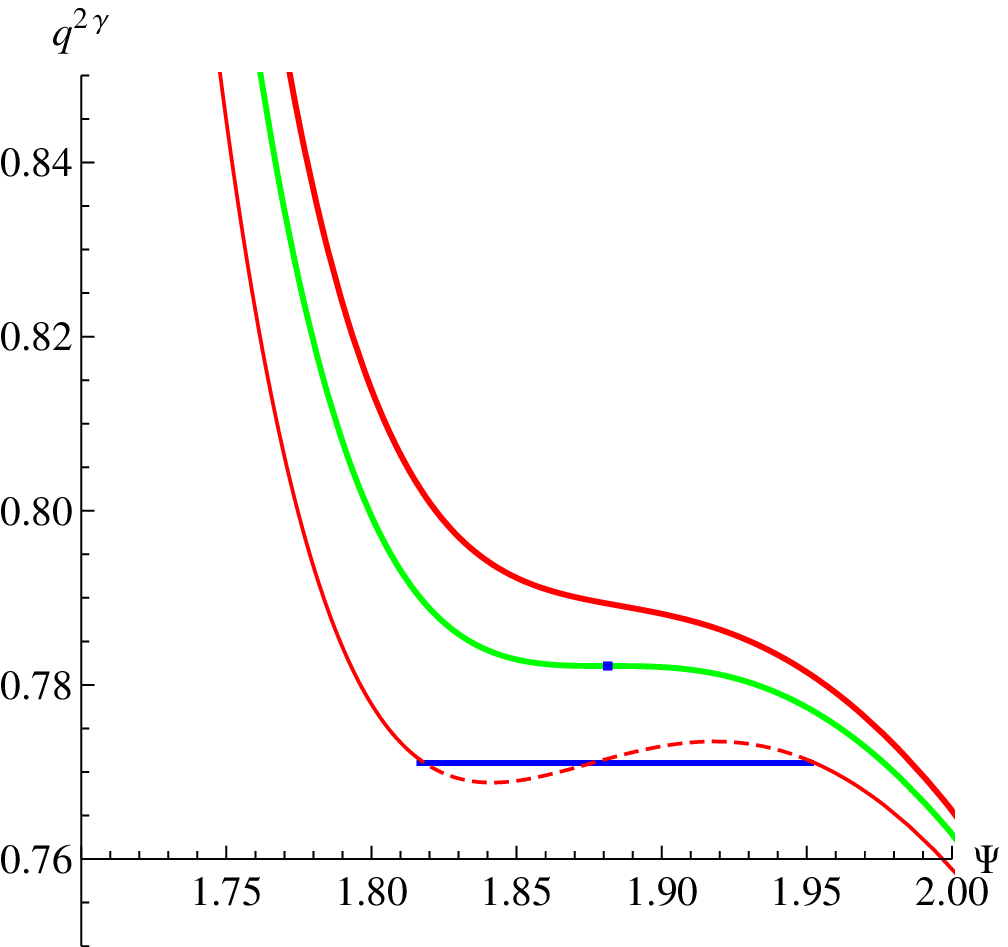}}
\caption{Phase transition curves in $q^{2\gamma}-\Psi$ diagram with the parameter $\gamma=0.8$. In the left panel, $\gamma=0.8,~P_0=0.0034$, and the temperatures is set to $T_0=0.0419$ (red thin line), $T_0=T_c=0.0417$ (green thick line), and $T_0=0.0415$ (red thick line). In the right panel, $\gamma=0.8,~T_0=0.0419$, and the pressures is set to $P_0=0.0034$ (red thin line), $P_0=P_c=0.003476$ (green thick line), and $P_0=0.0035$ (red thick line).}\label{Q-Psi}
\end{figure}

For the given temperature $T_0$ (i.e., $\chi$), the phase transition condition for the independent dual parameters $q^{2\gamma}-\Psi$ is consistent with that for $P-V$ and $T-S$. For the EPYM black hole with the fixed YM charge and temperature, when the horizon radius $r_+$ is smaller than $r_1$, the phase corresponds to the liquid of a van der Waals system; in contrast, it resembles the gas of a van der Waals system for $r_+>r_2$. And the phase corresponds to the two-phase coexistent of a van der Waals system as $r_1<r_+<r_2$. The phase transition curves for $q^{2\gamma}-\Psi$ with fixed temperature and pressure are shown in Fig. \ref{Q-Psi}, respectively.

\section{The coexistent curve in P-T diagram}
\label{scheme4}
For an ordinary thermodynamic system, when phases ($\alpha$ phase and $\beta$ phase) are in the two-phase coexistence area, the coexistent curve ($P-T$) is directly determined by experiments. The slope of the $P-T$ curve is given by the Clapeyron equation as
\begin{eqnarray}
\frac{d P}{d T}=\frac{L}{T\left(\nu^{\beta}-\nu^{\alpha}\right)}, \label{LPT}
\end{eqnarray}
where $L=T(S^\beta-S^\alpha)$; $\nu^{\alpha},~\nu^{\beta}$ are the molar volumes of $\alpha$ and $\beta$ phases, respectively. Generally, for an ordinary thermodynamic system, the Clapeyron equation agrees with the experimental results, which provide a direct experimental verification for the correctness of thermodynamics.

\begin{figure}[htp]
\centering
\includegraphics[width=0.45\textwidth]{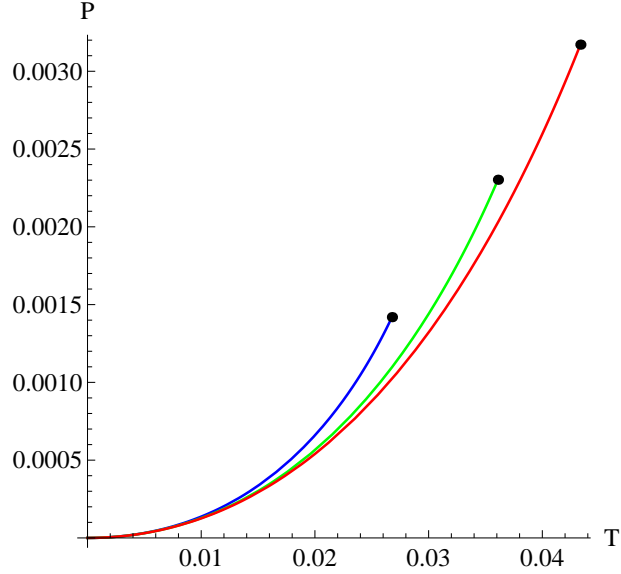}
\caption{Coexistent curve in $P-T$ diagram with the parameter $q=1.2$. The critical points with different values of the non-linear YM charge parameter $\gamma$ are marked as black dots.}\label{P-T}
\end{figure}

\begin{figure}[htp]
\centering
\includegraphics[width=0.45\textwidth]{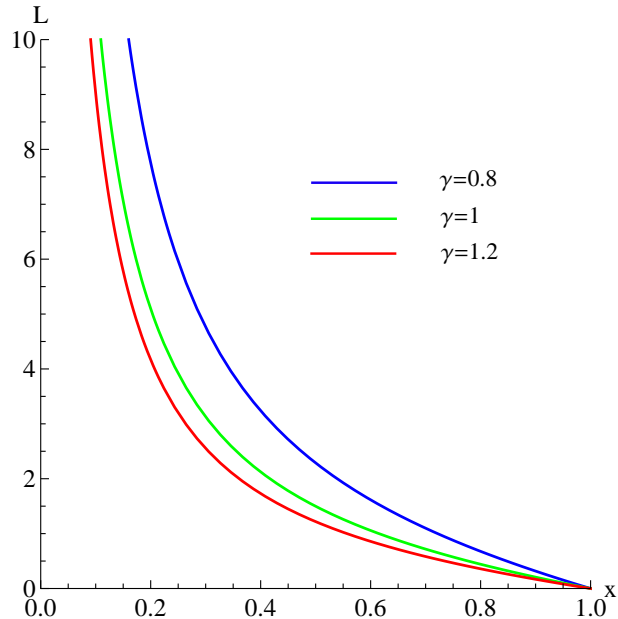}
\caption{$L-x$ curve with the parameter $q=1.2$. }\label{L-x}
\end{figure}

For the EPYM black hole, from Eqs. (\ref{T0}), (\ref{PPP0}) and (\ref{r2}), we have
\begin{eqnarray}
T&=&\frac{1}{4 \pi x\left(2 q^{2}\right)^{\gamma /(4 \gamma-2)} f^{1 /(4 \gamma-2)}(x, \gamma)}\nonumber\\
&&\times\left(1+x-\frac{1}{2 f(x, \gamma) x^{4 \gamma-2}} \frac{1-x^{4 \gamma}}{1-x}\right)=y_{2}(x, \gamma),~~~~  \label{LT}\\
P&=&\frac{1}{8 \pi r_{2}^{2} x}\left(1-\frac{\left(2 q^{2}\right)^{\gamma}\left(1-x^{4 \gamma-1}\right)}{2 r_{2}^{4 \gamma-2} x^{4 \gamma-2}(1-x)}\right)\nonumber\\
&=&\frac{1}{8 \pi x\left(2 q^{2}\right)^{\gamma /(2 \gamma-1)} f^{1 /(2 \gamma-1)}(x, \gamma)}\nonumber\\
&&\times\left(1-\frac{1-x^{4 \gamma-1}}{2 x^{4 \gamma-2}(1-x) f(x, \gamma)}\right)=y_{1}(x, \gamma).\label{LP}
\end{eqnarray}
The coexistent curves ($P-T$) with different non-linearity parameters $\gamma$ are shown in Fig. \ref{P-T}. Furthermore, from eqs. (\ref{LPT}), (\ref{LT}), and (\ref{LP}), the latent heat of phase transition for EPYM black hole reads
\begin{eqnarray}
L&=&T\left(1-x^{3}\right) r_{2}^{3} \frac{4 \pi y_{1}^{\prime}(x, \gamma)}{3 y_{2}^{\prime}(x, \gamma)}\nonumber\\
&=&\left(1-x^{3}\right) \frac{4 \pi y_{1}^{\prime}(x, \gamma)}{3 y_{2}^{\prime}(x, \gamma)}\left(2 q^{2}\right)^{3 \gamma /(4 \gamma-2)} f^{3 /(4 \gamma-2)}(x, \gamma) y_{2}(x, \gamma).
\end{eqnarray}
The above equation suggests that for the system with a fixed YM charge, the latent heat of phase transition is related with the temperature $T_0$ ($T_0\leq T_c$), that is, related with $x$. The latent heat of phase transition with $x$ for non-linear parameters $\gamma$ are shown in Fig. \ref{L-x}. The latent heat of phase transition for fixed $x$ decreases with an increase in $\gamma$ ($\gamma>\frac{3}{4}$).

\section{Thermodynamic geometry}
\label{scheme5}
Eq. (\ref{r2}) demonstrates that for the EPYM black hole with $T_0$ ($T_0\leq T_c$), there is a sudden change of the YM electric potential of the black hole molecules. The YM electric potentials with different phases are
\begin{eqnarray}
\phi_{2}^{2}=\frac{\left(2 q^{2}\right)^{\gamma}}{r_{2}^{4 \gamma-2}}=\frac{1}{f(x, \gamma)}, \qquad \phi_{1}^{2}=\frac{\left(2 q^{2}\right)^{\gamma}}{r_{1}^{4 \gamma-2}}=\frac{1}{x^{4 \gamma-2} f(x, \gamma)}.\label{phases}
\end{eqnarray}
This indicates that the microstructures of the black hole molecules in different phases, are inconsistent. Recent studies proposed that the phase transition of a black hole is due to the different number densities of molecules for large and small black holes \cite{Miao2018,Miao2019,Miao2017,Altamirano2014}. According to our investigation in the last section, the phase transition of the EPYM black hole is related with the horizon and the YM charge. That is, the phase transition is determined by the YM electric potential at the horizon. Based on this issue and Landau's theory of continuous phase transition, we explore to explore the physical mechanism of phase transition for the EPYM black hole.

The continuous phase transition theory is characterized by the change of the degree of order and the accompanying change of the symmetry property of matter. So what is the internal reason for the phase transition of the EPYM black hole? With the above investigation, the phase transition of the EPYM black hole with $T_0$ ($T_0<T_c$) occurs when the YM electric potential at the horizon satisfies the relation (\ref{PTC}). For the EPYM black hole with $T_0<T_c$ and phase $\phi_1^2$, the YM electric potential is high; hence, the black hole molecules are affected by this strong YM electric potential; consequently, they have a certain orientation and displacement polarization. In contrast, for this system with the same temperature $T_0<T_c$ and another phase $\phi_2^2$, the black hole molecules become disordered; these molecules have a higher symmetry due to a lower YM electric potential. The thermal motion of black hole molecules weakes the order of black hole molecules with increasing temperature. Especially for $T_0>T_c$, the thermal motion of black hole molecules leads to the disorder of black hole molecules. In other words, for the EPYM black hole with the lower temperature $T_0<T_c$, the black hole molecules have a lower symmetry and higher order, and the order parameter $\phi^2(T)$ is not equal to zero. While the black hole molecules have a higher symmetry and lower order, and the order parameter $\phi^2(T)$ is zero for the black hole with $T_0>T_c$. Note that the above results are obtained by comparing the EPYM black hole to an ordinary thermodynamic system.

\begin{figure}[htp]
\centering
\includegraphics[width=0.45\textwidth]{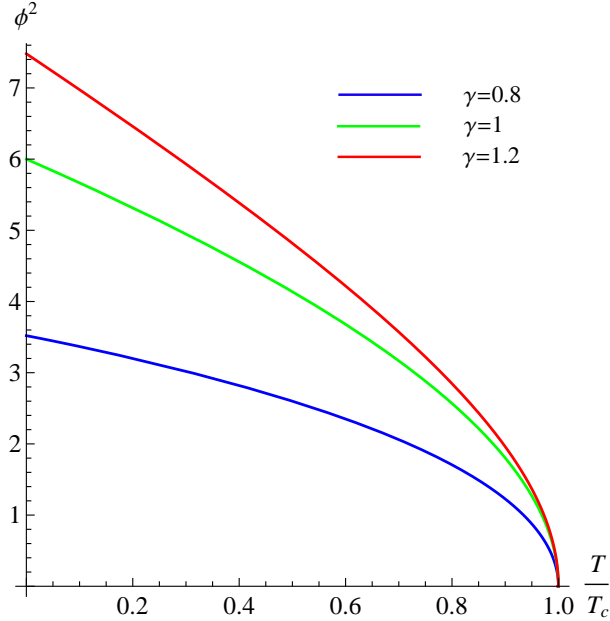}
\caption{$\phi^2$-$\frac{T}{T_c}$ curves for the EPYM black hole with the non-linear YM charge parameter $\gamma$.}\label{phi-x}
\end{figure}

We define the order parameter of the EPYM black hole as
\begin{eqnarray}
\phi^{2}(T)=\frac{\phi_{1}^{2}-\phi_{2}^{2}}{\phi_{c}^{2}}=\frac{\gamma(4 \gamma-1)( 1-x^{4 \gamma-2})}{f(x, \gamma) x^{4 \gamma-2}}
\end{eqnarray}
with $\phi^{2}_c=\frac{1}{f(1, \gamma)}=\frac{1}{\gamma(4 \gamma-1)}$. The order-parameter plotted with the non-linear parameter $\gamma$ is shown in Fig. \ref{phi-x}.

Becasue the order parameter is small near the critical point, according to Landau's theory of continuous phase transition, Gibbs function can be expanded as the power of $\phi^2(T)$ near the critical temperature $T_c$ by the method in Refs. \cite{Guo2020,Du2019}. The free energy takes the minimum value in the equilibrium state; based on this theory, we obtained that the critical exponents of the EPYM black hole are consistent with that of the RN-AdS black hole and the single axis ferromagnetic. However, the fluctuation of the order parameter near the critical point is neglected in the above discussion. Fortunately, the Ruppeiner geometry is derived from the theory of thermodynamic fluctuation. The singularity of the scalar curvature reveals the phase transition structure of black holes \cite{Ruppeiner2018,Ruppeiner1995}. In the following, we explore the microstructure of the black hole molecules by investigating the Ruppeiner geometry. We take ($S$, $P$) fluctuation with the fixed YM charge. The Ruppeiner scalar curvature (Ricci scalar) reads \cite{Miao2018}
\begin{eqnarray}
R=-\frac{N}{D}
\end{eqnarray}
with
\begin{eqnarray}
N&=&2 \pi S^{-1}[2^{1+\gamma}\pi^{1+2\gamma}q^{2\gamma} S^{1-2\gamma}(-1+2\gamma)[-1-48 P S\nonumber\\
&&+(-1+8 P S(7+16 P S))\gamma+2(1-8 P S)^{2}\gamma^{2}]\nonumber\\
&&+8\pi^{2}(-1+2 \gamma)[(-1+\gamma)^{2}+256 P^3S^3 \gamma(-1+2 \gamma)\nonumber\\
&&+4P S(1+\gamma-2 \gamma^{2})-32P^{2}S^{2}(-3+2 \gamma(2+\gamma))]\nonumber\\
&&+\pi^{4 \gamma}\left(2 q^{2}\right)^{2 \gamma} S^{2-4 \gamma}[-3+\gamma(9-6 \gamma+8 P S(1+2 \gamma))]],\\
D&=&\left[-\pi^{2 \gamma} \frac{\left(2 q^{2}\right)^{\gamma}}{S^{2 \gamma-1}}+2 \pi(1+8 P S)\right]\nonumber\\
&&\times\left[\pi^{2 \gamma} \frac{\left(2 q^{2}\right)^{\gamma}}{S^{2 \gamma-1}}+2 \pi(1-8 P S)(1-2 \gamma)\right]^{2}.
\end{eqnarray}
Because the forms of the horizon radius with a given temperature $T<T_c$ for two phases are different, the Ricci scalar of two phases also have two forms
\begin{eqnarray}
R_1&=&\nonumber\\
&&-\frac{2 (2q^2)^{\frac{1}{1-2\gamma}}f^{\frac{1}{1-2\gamma}}(x,\gamma)}
{\pi\left[-x^{2-4\gamma}f^{-1}(x,\gamma)+2(1+8A_1)\right]
\left[x^{2-4\gamma}f^{-1}(x,\gamma)+2(1-2\gamma)(1-8A_1)\right]^2}\nonumber\\
&&\times B,~~~~~~\\
R_2&=&-\frac{2 (2q^2)^{\frac{1}{1-2\gamma}}f^{\frac{1}{1-2\gamma}}(x,\gamma)}
{\pi\left[-f^{-1}(x,\gamma)+2(1+8A_2)\right]
\left[f^{-1}(x,\gamma)+2(1-2\gamma)(1-8A_2)\right]^2}\nonumber\\
&&\times C~~~~~~
\end{eqnarray}
with the following functions
\begin{eqnarray}
B&=&8x^{-2}(-1+2\gamma)\left[
(-1+\gamma)^2+256A_1^3\gamma(-1+2\gamma)+4A_1(1+\gamma-2\gamma^2)
-32A_1^2(-3+2\gamma(2+\gamma))\right]
\nonumber\\
&&+2x^{-4\gamma}f^{-1}(x,\gamma)(-1+2\gamma)\left[
-1-48A_1+8A_1\gamma(7+16A_1)-\gamma+2(1-8A_1)^2\gamma^2\right]\nonumber\\
&&+x^{2-8\gamma}f^{-2}(x,\gamma)\left[
-3+\gamma(9-6\gamma+8A_1(1+2\gamma))\right],\nonumber\\
C&=&8(-1+2\gamma)\left[
(-1+\gamma)^2+256A_2^3\gamma(-1+2\gamma)+4A_2(1+\gamma-2\gamma^2)
-32A_2^2(-3+2\gamma(2+\gamma))\right]
\nonumber\\
&&+2f^{-1}(x,\gamma)(-1+2\gamma)\left[
-1-48A_2+8A_2\gamma(7+16A_2)-\gamma+2(1-8A_2)^2\gamma^2\right]\nonumber\\
&&+f^{-2}(x,\gamma)\left[
-3+\gamma(9-6\gamma+8A_2(1+2\gamma))\right],\nonumber\\
A_1&=&\frac{x}{8}\left(1-\frac{x^{-4\gamma+2}-x}{2(1-x)f(x,\gamma)}
\right),~~~~~~
A_2=\frac{1}{8x}\left(1-\frac{x^{-4\gamma+2}-x}{2(1-x)f(x,\gamma)}
\right).\nonumber
\end{eqnarray}

\begin{figure}[htp]
\centering
\subfigure[$q=0.85,~1\geq\gamma>\frac{3}{4}$]{
\includegraphics[width=0.45\textwidth]{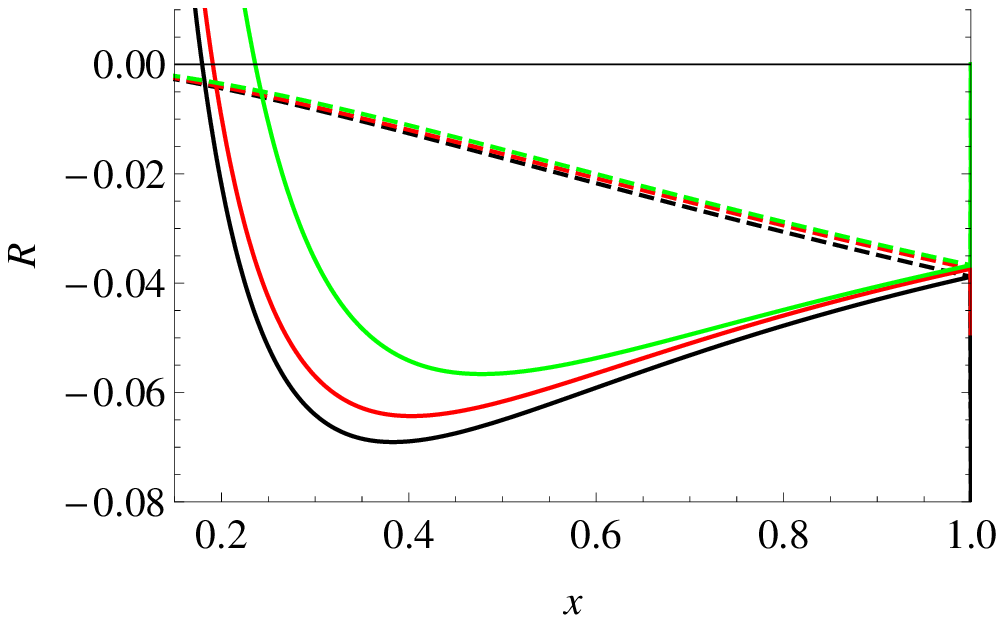}}~~~~
\subfigure[$q=0.85,~\gamma\geq1$]{
\includegraphics[width=0.45\textwidth]{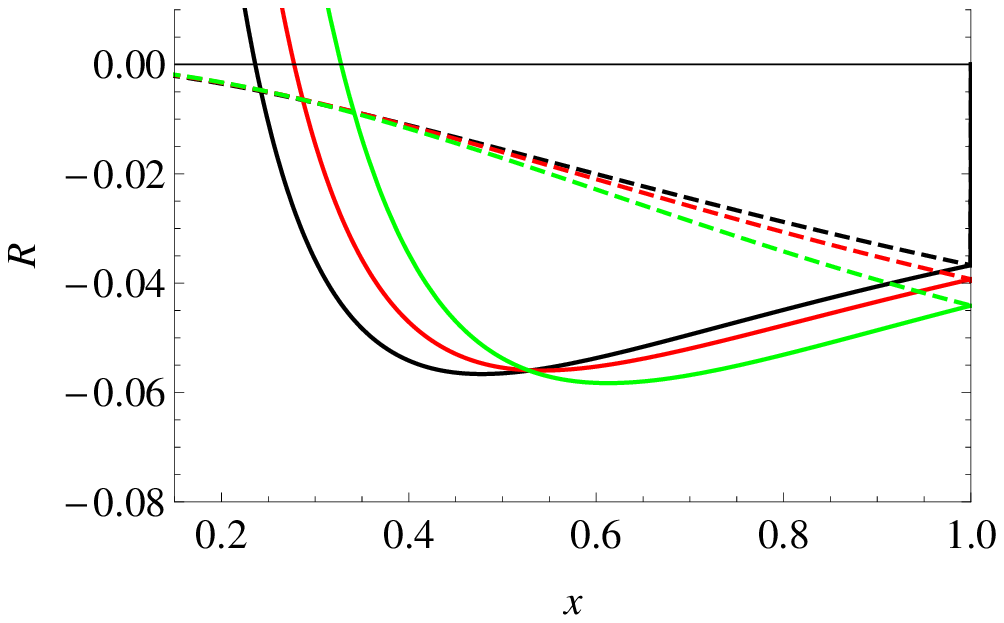}}~~~~\\
\subfigure[$\gamma=0.8$]{
\includegraphics[width=0.45\textwidth]{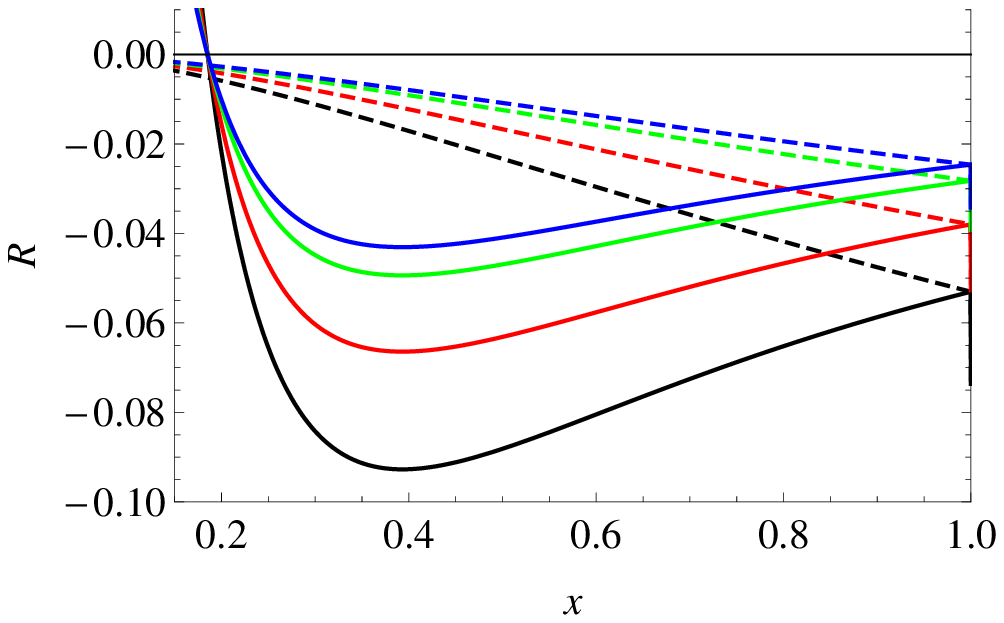}}
\caption{R-x curves with the different values of the non-linear parameter $\gamma$ and YM charge $q$ for two phases of the EPYM black hole. The dashed lines represent the geometry $R_2$, and the continuous lines correspond to the geometry $R_1$. In the upper left plot, $q=0.85,~1\geq\gamma>\frac{3}{4}$, and the non-linear parameter is set to $\gamma=0.78$ (black lines), $\gamma=0.82$ (red lines), and $\gamma=1$ (green lines). In the upper right plot, $q=0.85,~\gamma\geq1$, and the non-linear parameter is set to $\gamma=1$ (black lines), $\gamma=1.2$ (red lines), and $\gamma=1.5$ (green lines). In the bottom plot, $\gamma=0.8$, and the YM charge is set to $q=0.75$ (black lines), $q=0.85$ (red lines), $q=0.95$ (green lines), and $q=1$ (blue lines)}\label{R-x}
\end{figure}

Note that for the EPYM black hole with a given non-linear parameter $\gamma$ under going a phase transition, because of the conditions $P|_{r=r_1}\geq0$ and $\frac{\partial P}{\partial r}|_{r=r_1}\geq0$, the horizon radius $r_1$ has a minimum value, and $x$ must be from $x_{min}$ to one. In other words, $x_{min}$ must satisfy the following expression:  \begin{eqnarray}
\frac{(3-4\gamma)(1+x_{min})(1-x_{min}^{4\gamma})+8\gamma x_{min}^2(1-x_{min}^{4\gamma-3})}{2(3-4\gamma)(1-x_{min})^3}=\frac{4\gamma-1}{2}.
\end{eqnarray}
Especially, the minimum value of $x$ is independent of the YM charge $q$. The $R-x$ plots with the range $0\le x\leq1$ are shown in Fig. \ref{R-x}. For the given non-linear parameter $\gamma$, $x$ must be considered from some certain value, for example, $0.515\le x\leq 1$ with $\gamma=1.5$. From Fig. \ref{R-x}, we obtain that for the given YM charge $q$ and $x$, the distance ($\triangle R(x)=R_2(x)-R_1(x)$) of the Ricci scalars for two phases decreases with increaseint $\gamma$ and YM charge $q$. The effect of the non-linear parameter $\gamma$ on the Ricci scalar is consistent with that on the latent heat of phase transition given in the last section. It is considerably unique that the scalar curvature $R_c$ at the critical point increases with increasing YM charge $q$ and $\gamma$ in the range $1\geq\gamma>\frac{3}{4}$, while it decreases with increasing non-linear parameter $\gamma$ in the range $\gamma\geq1$.

Based on the conclusion of B. Mirza and H. Mohammadzadeh \cite{Mirza2008,Mirza2009} on the scalar curvature, the interactions of molecules are repulsive, attractive, and zero for $R>0$, $R<0$, and $R=0$, respectively. Fig. \ref{R-x} shows that for the EPYM black hole, $0>R_2>R_1$. That is, the average interaction of the black hole molecules for phase $\phi_2^2$ is less than that of phase $\phi_1^2$. Furthermore, the Ricci scalars for both phases increase with increasing of the YM charge $q$, until $R_2$ becomes zero. Since the density of black hole molecules $n=\frac{N}{V}=\frac{3}{\gamma l_p^2r_+}$, the density for phase $\phi_2^2$ is less than that for phase $\phi_1^2$. From Eq. (\ref{PTC}), we know that for the EPYM black hole with the fixed temperature ($T<T_c$), the phase $\phi_2^2$ corresponds to a lower YM electric potential, while the other phase $\phi_1^2$ corresponds to a higher YM electric potential. The horizon radius increases with increasing YM charge $q$, while the YM electric potential remains uncharged. Therefore, we think that the non-linear parameter $\gamma$ plays two roles in the phase transition for the EPYM black hole: change the order degree of the black hole molecules, which is determined by the YM electric potential, and change the density of the black hole molecules.

\section{Discussions and Conclusions}
\label{scheme6}
Because a detailed statistical description of the corresponding thermodynamic states of black holes is still unclear, the investigation of the thermodynamic properties and critical phenomenon of black holes becomes crucial. In this study, we focus on the phase transition and thermodynamic geometry of the four-dimensional Einstein-power-Yang-Mills (EPYM) black hole. We study the effect of the non-linear YM charge parameter $\gamma$ on the thermodynamical properties.

First, we reviewed the thermodynamic state parameters ($T,~P,~S,~M,~\Psi$) of the AdS black hole in the four-dimensional EPYM gravity. Then we reconstructed Maxwell's equal-area law by adopting different independent dual parameters to explore the phase transition of the EPYM black hole. The results showed that the phase transition point is the same for the three choices ($P-V,~T-S,~q^{2\gamma}-\Psi$). Especially, the independent dual parameters of the YM charge and the corresponding potential must be $q^{2\gamma}-\Psi$, not other forms. This result is attributed to the fact the phase transition must be independent with the concrete physical process. Furthermore, the phase transition condition of the EPYM black hole was presented as described in Eq. (\ref{PTC}). We obtain that the pressure of the EPYM black hole with the unchanged temperature $T\leq T_c$ decreases with increasing $\gamma$. The phase transition is related to the non-linear parameter $\gamma$ and the horizon radius, while it is not the only pure one between a small and a big black hole.

Next, we presented the coexistent curve in the $P-V$ diagram and analyzed the latent heat of phase transition for the EPYM black hole. The coexistent curve in the $P-T$ diagram showed that the critical point of phase transition increases with increasing non-linear YM charge parameter $\gamma$. In addition, the latent heat of phase transition for a fixed $x$ decreases with increasing non-linear parameter $\gamma ~(\gamma >3/4).$

Finally, by defining a new order parameter $\phi^2(T)$ and comparing the EPYM black hole to an ordinary thermodynamic system at the microscopic level, we explored the phase transition microstructure of the EPYM black hole by the scalar curvature $R$. The result suggested that the phase transition of the EPYM black hole is accompanied by the change of the black hole molecules' symmetry. For the EPYM black hole with a certain temperature and pressure ($T<T_c$, $P<P_c$), the Ricci scalars at the boundary of the two-phase coexistence area are different due to the different values of the order parameter $\phi^2$ at the boundary of the two-phase coexistence area. Especially, it was significantly unique that the scalar curvature $R$ at the critical point increases with increasing YM charge $q$ and the non-linear parameter $\gamma$ in the range $1 \geq \gamma >3/4$, while it decreases with increasing non-linear parameter $\gamma$ in the range $\gamma \geq 1$.

This work elucidated the phase transition microstructure of the EPYM black hole, which is useful to explore the microstructure of a black hole and understand its basic property. In particular, the in-depth study of the black hole microscopic structure will be significantly important to establish quantum gravity.

\section*{Acknowledgements}

We would like to thank Prof. Zong-Hong Zhu and Meng-Sen Ma for their indispensable discussions and comments. This work was supported by the National Natural Science Foundation of China (Grant No. 11705106, Grant No. 11475108, Grant No. 12075143), the Natural Science Foundation of Shanxi Province, China (Grant No.201901D111315), the Natural Science Foundation for Young Scientists of Shanxi Province, China (Grant No.201901D211441), the Scientific Innovation Foundation of the Higher Education Institutions of Shanxi Province (Grant Nos. 2020L0471, Grant Nos. 2020L0472), and the Science Technology Plan Project of Datong City, China (Grant Nos. 2020153).

\end{document}